\documentclass[aps, prb, reprint, superscriptaddress]{revtex4-2}
\usepackage{graphicx}
\usepackage{dcolumn}
\usepackage{bm}
\usepackage{amssymb}
\usepackage{amsmath}
\usepackage{epsf}
\usepackage{color}
\usepackage[colorlinks,citecolor=blue,linkcolor=blue]{hyperref}
\usepackage{titlesec}
\draft

\begin{document}

\preprint{APS/123-QED}

\title{Antiferromagnetic order and magnetic polarons in lightly doped Li$_x$CoO$_2$ (x $\sim$ 0.9)\\}

\author{Sudip Pal}
\email{sudip@csr.res.in}
\affiliation{UGC-DAE Consortium for Scientific Research, University Campus, Khandwa Road, Indore-452001, India}
\author{Pampa Sadhukhan}
\affiliation{Physikalisches Institut, Universität Würzburg, 97074 Würzburg, Germany}

\author{Waqar Suleman}
\affiliation{Institute of Solid State Physics, TU Wien, 1040 Vienna, Austria}
\author{Andrej Pustogow}
\affiliation{Institute of Solid State Physics, TU Wien, 1040 Vienna, Austria}
\author{S. B. Roy} 
\affiliation{UGC-DAE Consortium for Scientific Research, University Campus, Khandwa Road, Indore-452001, India}

\date{\today}
\begin{abstract}
We investigate the magnetic properties of Li$_x$CoO$_2$ (x$\sim$0.9) using bulk magnetization, specific heat, nuclear magnetic resonance (NMR) and electron paramagnetic resonance (EPR) spectroscopy measurements. The dc magnetization, specific heat and NMR measurements, which probe the macroscopic response, indeed show that this compound partially undergoes an antiferromagnetic transition below $T_N \sim$ 10 K. In addition, we observed a weak ferromagnetic response, which gives rise to the history dependence in magnetization measurements at low fields and is observed at temperatures above room temperature. We propose that there are ferromagnetic clusters at high temperatures due to the formation of magnetic polarons out of doped holes. In EPR measurements performed at the $X$-band frequency, only a fraction of the total spins contribute and show Curie-like paramagnetic behavior as reflected in the temperature dependence of the EPR intensity. The temperature variation of the EPR spectra can be understood in the framework of the diffusion of magnetic polarons. 
\end{abstract}                            
\maketitle

\section{\label{sec:level}Introduction:}
Li$_x$CoO$_2$ is the primary cathode material in rechargeable Li-ion batteries and hence an essential component of power sources in recent times. As a result, understanding the magnetic and electronic properties of this compound is becoming increasingly important. In addition to the technological interest, other phenomena like the metal-insulator transition, magnetic and charge ordering, anomalous thermoelectric effects and the discovery of superconductivity in sister compound Na$_x$CoO$_2$ provided additional thrust to this field {\color{blue}\cite{Marianetti2004,Schaak2003,Ikedo2010,Young2013,Bernhard2007,Balicas2008,Julien2008,Shang2022,Ola2025}}. 

Li$_x$CoO$_2$ forms a layered structure with Li residing between layers of edge-sharing CoO$_6$ octahedra alternatively stacked along the c-axis. In the CoO$_2$ layers, Co atoms are arranged in triangular lattice and are therefore prone to geometrical frustration. In the strictly stoichiometric compound (x = 1), Co is formally in the 3+ valence state. The crystal field splits the Co-3d states into a lower manifold containing three levels ($t_{2g}$ states) and a higher manifold containing two levels ($e_g$ states). Co$^{3+}$ exists in the low-spin $t_{2g}^6e_g^0$ configuration (S = 0), and the compound is a band insulator. In Li$_x$CoO$_2$ (x$<$1) Li-ion deficiency is expected to result into partial transformation of Co$^{3+}$  to Co$^{4+}$ with $t_{2g}^5e_g^0$ electronic configuration (low spin state, $S = 1/2$) and also leads to creation of holes in the $t_{2g}$ valence band {\color{blue}\cite{Czyzyk1992,Galakhov2006,Mizokawa2013}}. X-ray absorption and photo-electron spectroscopy however gives indication towards oxygen-hole in Li-deficient samples. Nevertheless, despite a large body of experimental works, the magnetic and electronic properties of Li$_x$CoO$_2$ are still a mystery {\color{blue}\cite{Mukai2007,Elp1991,Motohashi2009,Miyoshi2010,Mizokawa2013,Yang2012,Sugiyama2005}}. Primarily, it is not understood how Li vacancies affect magnetism and whether there is a long-range magnetic order in this compound {\color{blue}\cite{Uthayakumar2014,Kiyotaka2018}}. Experimental studies on Li$_x$CoO$_2$ (x$\leq$1), focusing on the magnetism of this compound, include diverse techniques such as the bulk magnetization measurements, muon spin resonance ($\mu$SR), electron paramagnetic resonance (EPR), nuclear magnetic resonance (NMR) spectroscopy and neutron scattering experiments. In spite of this intense research activity, experimental results are rather puzzling, yet interesting. Various possibilities have been proposed to explain these observations. For instance, temperature-dependent charge disproportionation (Co$^{3+}$$\rightarrow$ Co$^{2+}$ + Co$^{4+}$), spin-state transition, spin fluctuations, or surface effects, etc.{\color{blue}\cite{Kawasaki2009}}. Besides, the origin of the insulating behavior in lightly doped Li$_x$CoO$_2$ (0.9$<$x$<$1) is also currently discussed {\color{blue}\cite{Marianetti2004,Mizokawa2013,Ahn2023}}. In this doping range, the onsite coulomb interaction of holes might not be a dominant factor, as the double occupancy of holes at a single site is rare. Therefore, a Mott insulating state formed out of the impurity states has been suggested {\color{blue}\cite{Marianetti2004}}. On the other hand, the possibility of localized polaron formation has also been proposed {\color{blue}\cite{Mizokawa2013,Ahn2023}}. Nevertheless, the existence of the polarons is not yet conclusive in experiments. In addition, whether it has any significant effect on the magnetic properties of this compound is also not clear. 

In this work, we have prepared a polycrystalline sample of the battery material Li$_x$CoO$_2$ with a slight Li-deficiency (x $\sim$ 0.9) and investigated its magnetic property using bulk magnetization, heat capacity, NMR and EPR spectroscopy measurements. We show that a small fraction of the doped holes undergoes a magnetic transition to an antiferromagnetic state below $T_N$ $\sim$ 10 K. The antiferromagnetic-like transition is visible as a broad and weak hump-like feature in bulk magnetization and heat capacity measurements around $T_N$. However, at higher temperatures and low magnetic fields, we also observed signatures of history-dependent magnetization, which is likely a reminiscent of weak ferromagnetism. We suggest that the weak ferromagnetism arises due to the formation of bound magnetic polarons, which can exist as nanoscale inhomogeneities formed out of doped holes {\color{blue}\cite{Ogarkov2006,Kuhns2003,1Phelan2006,Kochelaev1997}}. The magnetic property of this compound is intricately connected to the diffusion of Li vacancies, and it explains the observations made in electron paramagnetic resonance spectroscopy.


\section{Experimental details} 
Polycrystalline sample of Li$_x$CoO$_2$ has been prepared by the conventional solid-state method. Stoichiometric amount of Li$_2$CO$_3$ and Co$_3$O$_4$ powders with a purity of 99.99\% is thoroughly mixed, followed by several hours of grinding. Then the mixture was heated at 800$^\circ$C for 24 hours with intermediate grindings to prepare Li$_x$CoO$_2$ polycrystalline samples. The phase formation of the sample has been checked at room temperature in lab-based Rigaku Rotaflex RTC300 RC diffractometer with 18 kW rotating anode Cu-K$_\alpha$ x-ray source and analyzed by rietveld refinement using Fullprof software. We have also characterized the powder sample in synchrotron (INDUS-2 at Raja Rammana Center for Advanced Technology, Indore) x-ray sources.  Magnetization measurements have been carried out in 7 Tesla superconducting quantum interference device magnetometer (MPMS3 SQUID-VSM, M/S Quantum Design, USA). Heat Capacity of a single piece of Li$_x$CoO$_2$ pallet weighing approximately 6 mg has been measured in the relaxation method using 9 Tesla Physical Property Measurement System (PPMS, M/S Quantum Design, USA). To determine the stoichiometry of the sample, we have used two different techniques, namely the ICP-OES (Inductively coupled plasma-atomic emission spectroscopy) to analyze Li and Co and the hot gas extraction method to determine oxygen content. The ICP-OES measurements were performed by using SPECTRO CIROS. 
Nuclear magnetic resonance (NMR) measurements were carried out on $^7Li$ nuclei at magnetic fields from 3 to 5.7~T in a Quantum Design PPMS cryostat at TU Wien covering a temperature range 2--100~K. The acquired spectra were analyzed at full recovery and the relaxation rate was determined by the saturation recovery method.
The EPR experiments in the temperature range between 5 and 300 K were performed using Bruker EMX $X$-band spectrometer at a frequency of $\nu=$ 9.47 GHz on a single piece of polycrystalline pallet using a continuous flow helium cryostat. The first derivative of the EPR absorption spectra ($dP/dH$, where $P$ is the absorbed microwave power) is recorded using a phase-sensitive lock-in amplifier detection technique. 

\begin{figure}[t]
	\centering
	\includegraphics[scale=0.3]{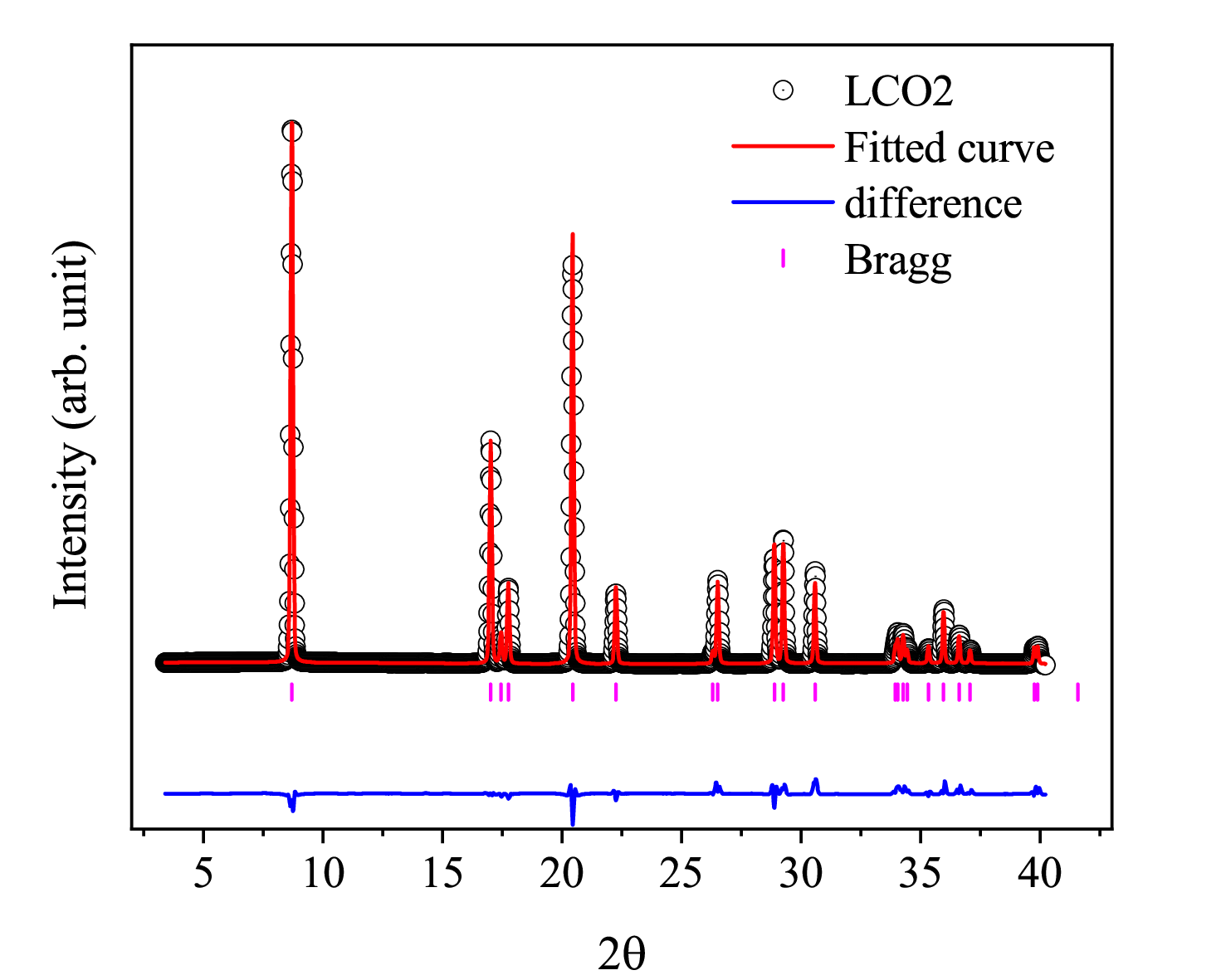}
	\caption{XRD pattern of the Li$_{0.9}$CoO$_2$ powder sample recorded at room temperature using synchrotron x-ray source of wavelength $\lambda$ = 0.7112 {\AA} (open circles). The red line shows the rietveld refinement of the powder spectra using Fullprof software. Blue line shows the difference between experimental data and the simulated pattern.}
\end{figure}



\begin{figure*}[t]
	\centering
        \includegraphics[scale=0.6]{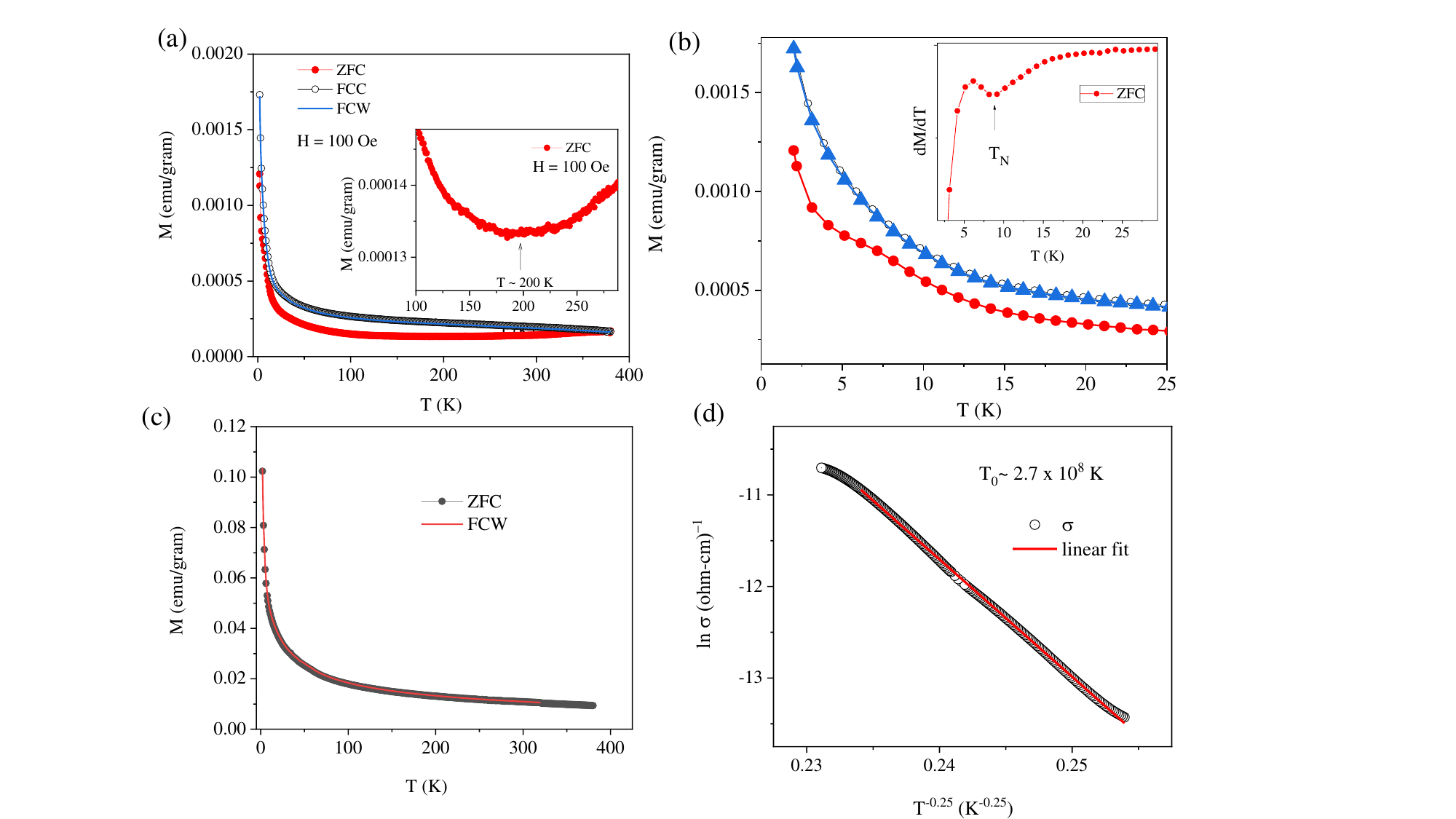}
	\caption{(a) $M-T$ curves measured at $\mu_0H$ = 100 Oe following the $ZFC$, $FCC$ and $FCW$ protocols. Inset shows the minima observed around 200 K in the $ZFC$ (b) the low temperature $M-T$ at $\mu_0H$ = 100 Oe. Inset shows the first derivative of the $ZFC$ to highlight the anomaly at $T_N$ (c) It shows  $M-T$ curve at $\mu_0H$ = 1 Tesla, (d) the semi-log plot of $\sigma$ vs. $T^{-0.25}$. The red line is the linear fit to the data.}
\end{figure*}

\section{Experimental Results}

\subsection{Sample details} 
In Fig. {\color{blue}1}, we present the XRD pattern of LiCoO$_2$ powder samples recorded at room temperature using a synchrotron x-ray source of wavelength $\lambda$ = 0.7112 {\AA}. All the observed reflections in the powder pattern can be nicely indexed by assuming a rhombohedral crystal structure with R${\bar3}$m space group without any detectable secondary phase. The Li, Co, and O atoms occupy the Wyckoff positions 3b, 3a, and 3c, respectively, where the z coordinate of the position parameter of the oxygen atom is refinable. The lattice parameters obtained from the Rietveld refinements of the powder pattern are a=b= 2.814 {\AA} and c= 14.051 {\AA}, which are similar to the reported values in the literature. The composition of the LiCoO$_2$ with the general formula Li$_x$Co$_y$O$_{z}$ is found to be x = 0.92(0.03), y = 1.0, and z = 2.04(0.02). The lower value of Li than expected (x = 1) is most likely due to evaporation of Li during sample preparation {\color{blue}\cite{Yang2012}}.

\subsection{Magnetization and resistivity}
In Fig. {\color{blue}2(a)}, we show the magnetization as a function of temperature ($M-T$) curve at $\mu_0H$ = 100 Oe measured in the zero field cooling ($ZFC$), field cooled cooling ($FCC$) and field cooled warming ($FCW$) protocols. In the $ZFC$ protocol, the sample is initially cooled down to the lowest possible temperature in the absence of any magnetic field. At the lowest temperature, $\mu_0H$ = 100 Oe is applied and $M-T$ has been recorded during warming. The $FCC$ and $FCW$ curves have been measured during subsequent cooling and heating cycles without changing the magnetic field. The temperature sweep rates during the initial cooling and the measurements are 10 K/min and 1.5 K/min, respectively. All the curves are weakly dependent of temperature down to 50 K and rapidly increase at further lower temperatures. Such a temperature dependence is quite similar to previous reports where the $M-T$ curves were measured at higher field {\color{blue}\cite{Mukai2007}}. However,  in the $M-T$ measurement in Fig. {\color{blue}2(a)}, which is recorded at 100 Oe, we can make some interesting and non-trivial observations that have not been highlighted before:

(i) The most interesting observation is a finite thermomagnetic irreversibility between the $ZFC$ and $FCW$/$FCC$ curves. The irreversibility gradually decreases with increasing temperature but persists till the maximum temperature of the measurements, which is $T \sim$ 380 K. There is no difference between the $FCC$ and $FCW$ curves. The bifurcation between the $ZFC$ and $FCW$ curves gradually suppresses with an increase in the fields. At $\mu_0H \sim$ 1 Tesla (Fig. {\color{blue}2(c)}) two curves are identical as reported in the literature.

(ii) A close look at the $ZFC$ curve further reveals that as we reduce temperature, the $ZFC$ magnetization actually decreases slowly. It exhibits a broad minimum with the lowest magnetization around $T \sim$ 200 K (inset of Fig. {\color{blue}2(a)}), and increases again rather sharply at even lower temperatures. Whereas the $FCC/FCW$ curves monotonically increase down to the lowest temperature of the measurement.

(iii) We see a weak and broad hump around $T_N$ $\sim$ 10 K more clearly visible in the $ZFC$ curve (main panel of Fig. {\color{blue}2(a)}), which resembles an antiferromagnetic transition. This becomes more prominent in the derivative curve (inset of Fig. {\color{blue}2(a)}).

(iv) We attempted to determine the equilibrium state of the system. To do that, we performed thermal cycle experiments both in the $ZFC$ and $FC$ states as shown in Fig. {\color{blue}3(a,b)} {\color{blue}\cite{Roy2007,Sudip2020, Sudip2021, Sudip2025}}. We observed that only in the $ZFC$ state, the magnetization monotonically increases after each thermal cycle. This indicates the non-equilibrium character of the $ZFC$ state.

 \begin{figure}[t]
	\centering
	\includegraphics[scale=0.45]{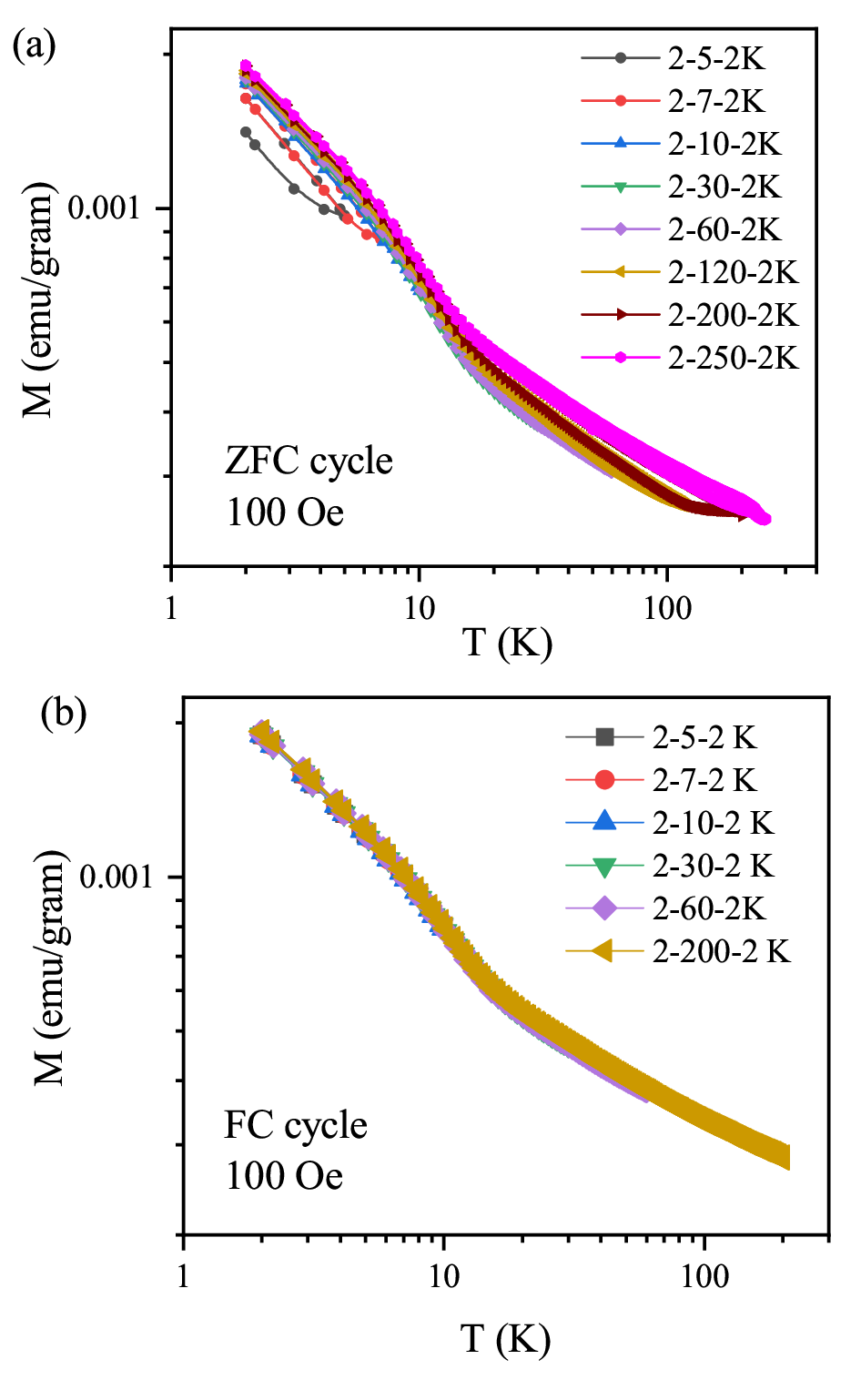}
	\caption{The double-log plot of the $M-T$ curves recorded following thermal cycles from progressively higher temperatures measured at $\mu_0H$ = 100 Oe following the (a) $ZFC$ protocol. (b) $FC$ protocol.}
\end{figure}

\begin{figure}[t]
	\centering
	\includegraphics[scale=0.42]{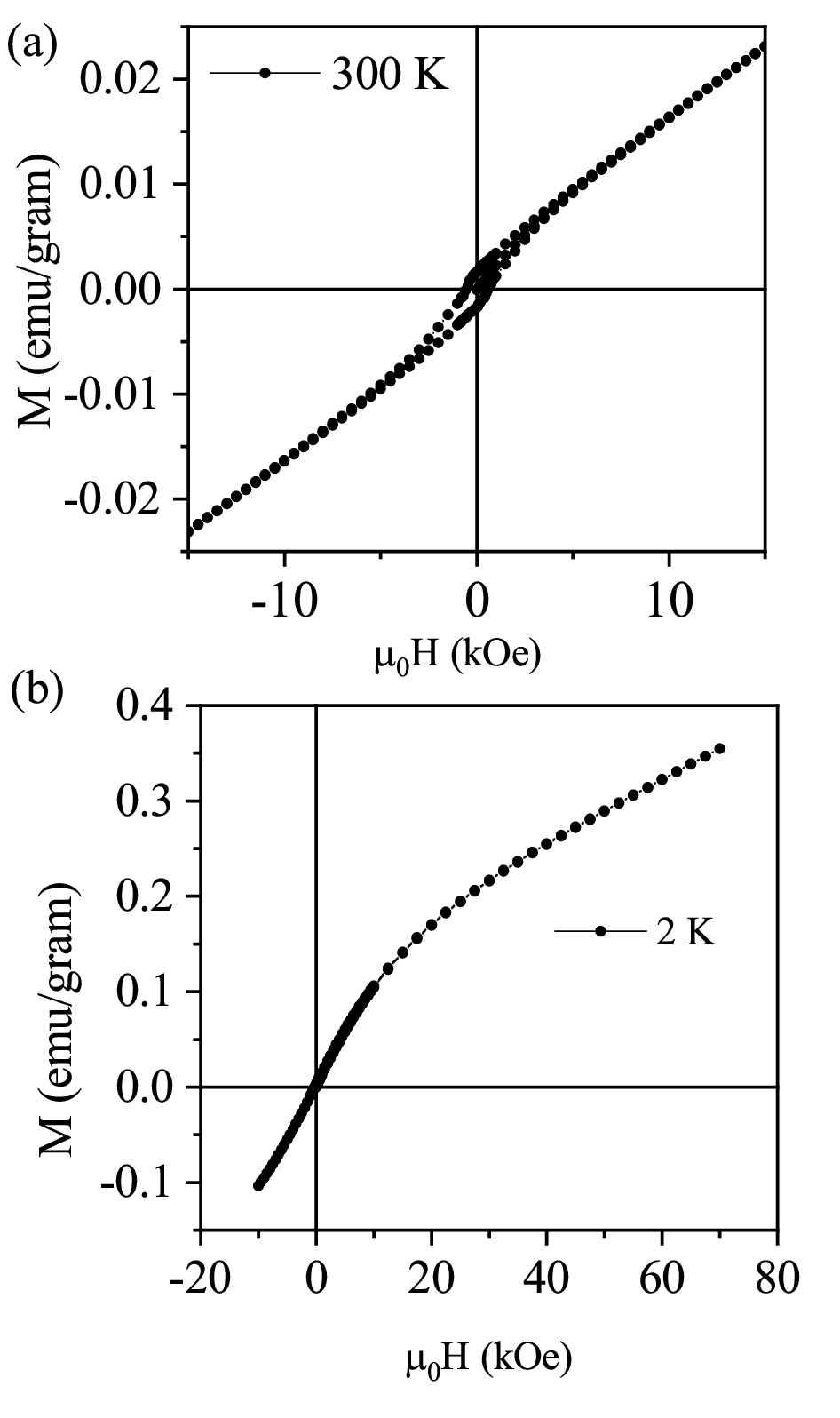}
	\caption{Isothermal $M-\mu_0H$ curve at (a) $T$ = 300 K and (b) 2 K.}
\end{figure}

In addition to the history-dependent magnetization, interestingly, the isothermal magnetization measurements as a function of magnetic field ($M-\mu_0H$ curve) at room temperature as shown in Fig. {\color{blue}4(a)} clearly show a hysteresis at the origin. This also indicates that the magnetic state at room temperature is not purely paramagnetic and points towards the presence of short-range ferromagnetic interaction. On the other hand, the $M-\mu_0H$ increases nearly linearly after the closure of the hysteresis loop. Such a nearly linear increase in $M-\mu_0H$ at higher field indicates the co-existence of isolated spins showing paramagnetic response along with the weak ferromagnetism. At 2 K, the $M-\mu_0H$ as shown in Fig. {\color{blue}4(b)}, also shows similar behavior, however, with reduced coercivity.

In this context, it may be noted that in a previous investigation, signatures of history dependence in bulk magnetization were observed in a similar temperature range {\color{blue}\cite{Miyoshi2010}}. The two $ZFC$-$M(T)$ curves, when measured at different cooling rates, were observed to bifurcate below a certain composition-dependent temperature. Interestingly, in a similar temperature range, the Li-NMR studies showed temperature dependence of the spin-lattice relaxation time, $T_1$, and the line width of the $^7$Li nucleus, which is due to the reduction of Li-ion diffusion with reducing temperatures {\color{blue}\cite{Nakamura2006}}. Therefore, the history dependence observed in magnetization measurements is suggested to originate from the freezing of the Li-ion diffusion. However, it is not clear how the freezing of the Li-ion diffusion can induce such an effect on the bulk magnetization. Here, we propose that the bifurcation between the $ZFC$ and the $FCW$ $M(T)$ curves at low magnetic fields as observed in our experiments, is related to the formation of magnetic polarons.     

In systems with localized carrier states, the variable range hopping (VRH) of polarons is a possible candidate for explaining the conduction process. In general, the conductivity using Mott's VRH can be expressed as $\sigma = \sigma_0exp(T_0/T)^{-p}$, where $p = 1/(d+1)$, $d$ being the dimensionality of the system, $T_0$ is Mott's activation energy (in units of K). In our sample, the transport data in the measurement temperature range can be fitted to Mott's formula for 3-dimensions, with $p=1/4$. A semi-log plot of $\sigma$ vs T$^{-0.25}$ gives a straight line as shown in Fig. {\color{blue}2(d)}.

\begin{figure}[t]
	\centering
	\includegraphics[scale=0.52]{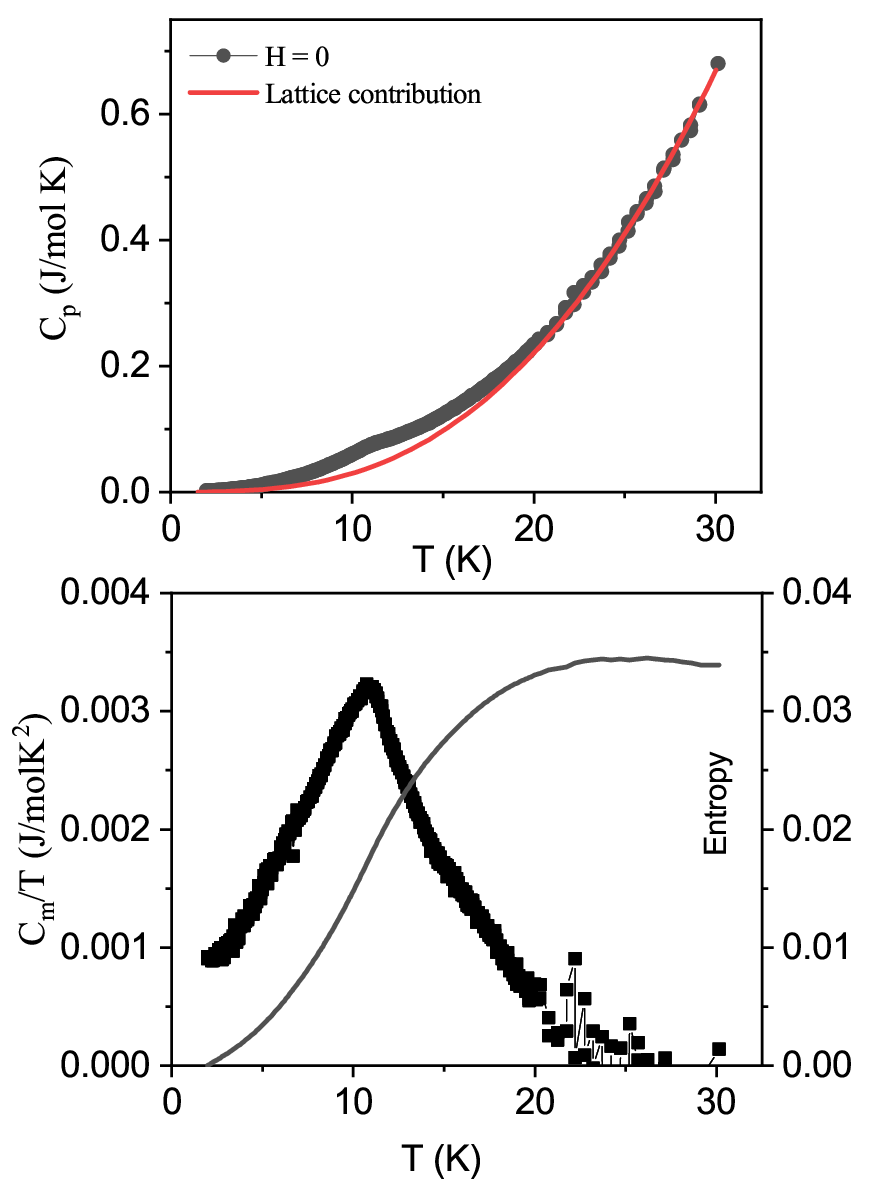}
	\caption{(a) The temperature variation of heat capacity at H = 0. The red line is the lattice contribution to the total heat capacity. (b) shows the magnetic contribution to the heat capacity around the magnetic transition.}
\end{figure}

\begin{figure}[h]
	\centering
	\includegraphics[scale=0.36]{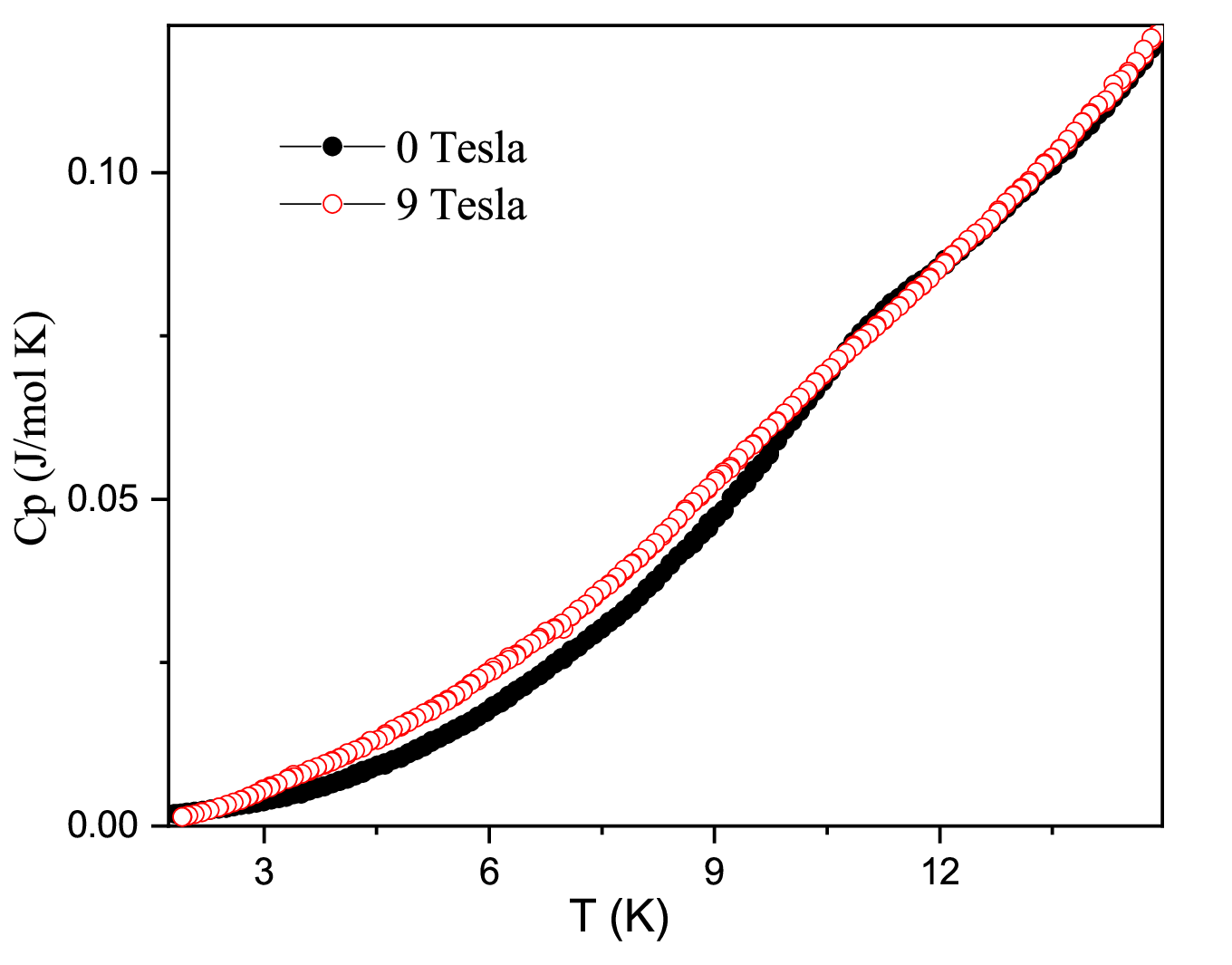}
	\caption{Magnetic field dependence of the heat capacity. At higher fields, the anomaly becomes broader and shifts towards lower temperatures.}
\end{figure}


\subsection{Heat capacity:}
In Fig. {\color{blue}5(a)}, we show the heat capacity ($C_P$) of Li$_{0.9}$CoO$_2$ as a function of temperature up to $T =$ 30 K in absence of an external magnetic field. Around $T \sim$ 10 K, a weak and broad $\lambda$-like anomaly is observed, which is likely associated with the antiferromagnetic transition, consistent with the magnetization measurements. We have also shown the lattice contribution to the specific heat as a red continuous line. To estimate the lattice contribution, we have fitted the total heat capacity in the temperature range between $T\sim$ 20-50 K by a polynomial equation given by {\color{blue}\cite{Pal2024}} $	C_{ph} = \beta T^3 + \gamma T^5$ and then extended down to $T =$ 2 K.

The magnetic contribution to heat capacity across the transition can be found by subtracting $C_{ph}$ from the total heat capacity, which we show in Fig. {\color{blue}5(b)}. The magnetic entropy released across the phase transition is only 0.035 K/mole K, which is only 7\% of total hole-doping resulting from the estimated Li-deficiency. The absence of a sharp specific heat anomaly and the small entropy  indicates that only a very small fraction of the doped holes forms antiferromagnetically ordered clusters below $T\sim$ 10 K. This is also consistent with the temperature dependence of magnetization where a weak and broad hump-like anomaly is observed near $T\sim$ 10 K.

To further study the anomaly at $T_N$, we have measured heat capacity at 9 Tesla. In Fig. {\color{blue}6} we compare the $C_P$ vs. $T$ of Li$_{0.9}$CoO$_2$ in the vicinity of the magnetic transition at zero and 9 Tesla magnetic fields. Note that at 9 Tesla, the $\lambda$-like feature in the heat capacity further broadens, and the maxima of the peak position is slightly shifted towards lower temperatures. This is in line with the conventional antiferromagnetic state where the external magnetic field of a few Tesla suppresses the heat capacity anomaly by a few degrees {\color{blue}\cite{Yang1995}}. 

\subsection{Nuclear magnetic resonance}
\begin{figure}[h]
	\centering
	\includegraphics[scale=0.45]{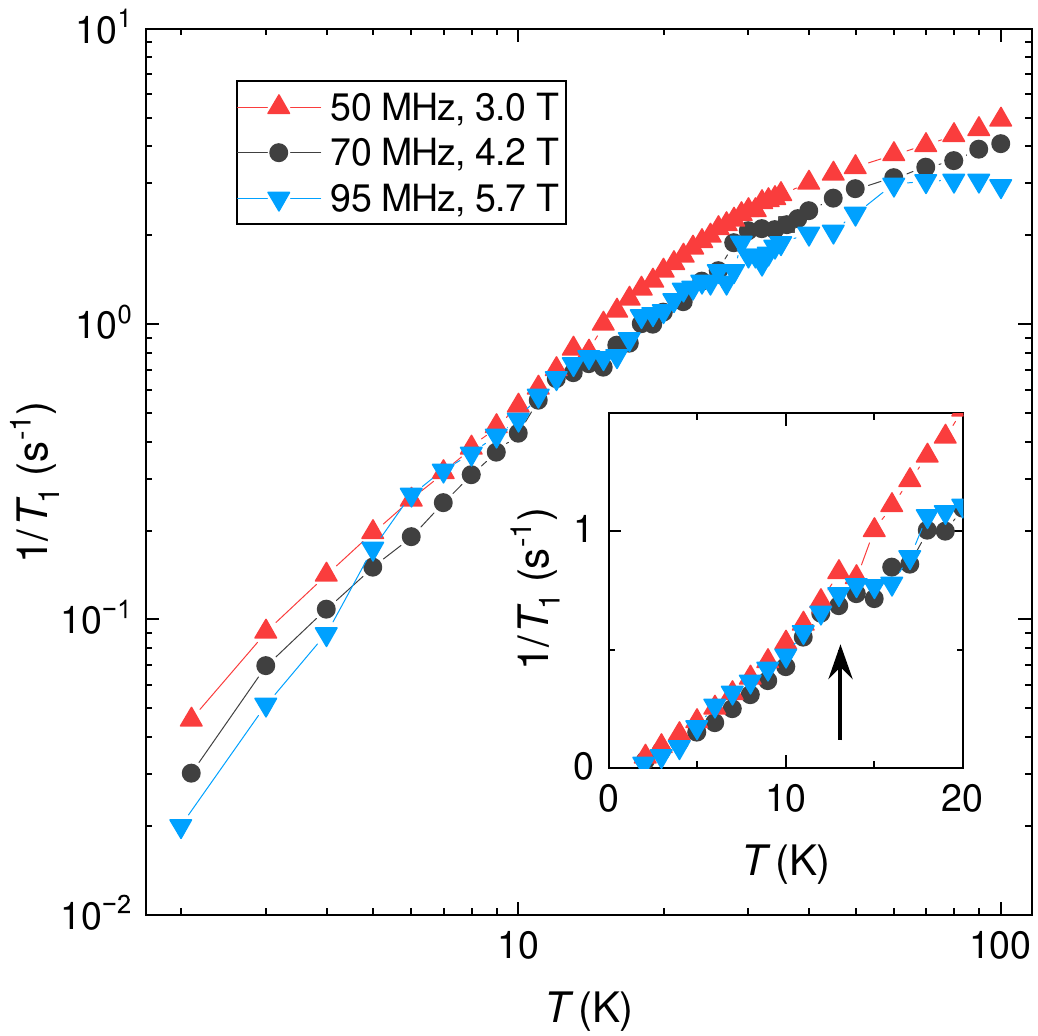}
	\caption{The $^7$Li NMR relaxation rate was measured at 3.0, 4.2 and 5.7~T. Inset: the bump slightly above 10 K indicates antiferromagnetic order.}
\label{NMR-T1}
\end{figure}

\begin{figure}[h]
	\centering
	\includegraphics[scale=0.35]{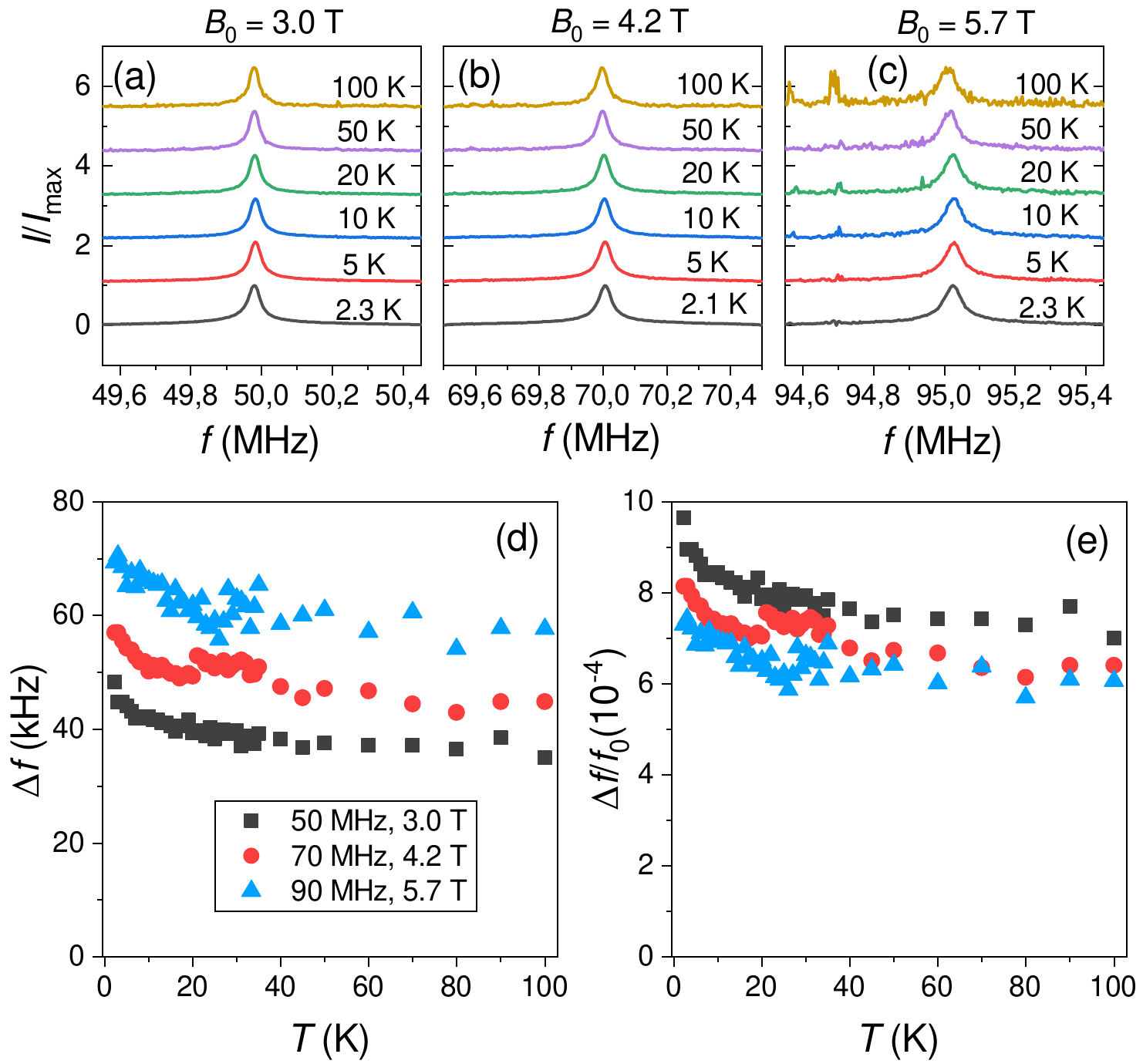}
	\caption{(a-c) $^7$Li NMR spectra at 3.0, 4.2 and 5.7~T yield a single line at all temperatures. (d) The linewidth (FWHM) exhibits a poor temperature dependence above 40K and increases below that, with a kink around 10~K. (e) The relative linewidth is rather similar for all fields, indicating paramagnetic broadening.}
\label{NMR-linewidth}
\end{figure}

In Fig.~\ref{NMR-T1}, we plot the NMR spin-lattice Relaxation rate $T_1^{-1}$ as a function of temperature on double-logarithmic scales. The log-log plot yields an exponent around $T^2$ at lowest temperatures, and its steepness reduces at higher temperatures, being sublinear above 40~K. As highlighted in the inset, there is a bump slightly above 10~K, which roughly coincides with the antiferromagnetic order deduced from magnetization and specific heat measurements.  We also determined the linewidth from the NMR spectra at all temperatures, which are displayed in Fig.~\ref{NMR-linewidth}. The line shape is a single resonance at all fields and temperatures, reflecting that the spin density around $^7$Li nuclei is small and does not vary largely as the unpaired valence electrons reside primarily on the Co ions. This is also the reason why no prominent line splitting is observed upon cooling through 10~K, as would be expected for an AFM transition. Still, a noticeable increase in line width confirms the antiferromagnetic transition, both on absolute (Fig.~\ref{NMR-linewidth}d) and relative scales (Fig.~\ref{NMR-linewidth}e).
Besides the broadening of the antiferromagnetic transition in $T_1^{-1}$ similar to the specific heat results and a general slowing down of relaxation at above 20~K, there is no significant field dependence in the inspected field range from 3.0 to 5.7~T, neither in relaxation rate nor in the NMR linewidth. Normalizing the line width to the resonance frequency in Fig.~\ref{NMR-linewidth}(e) yields collapse of the data, indicating predominant paramagnetic line broadening.

\begin{figure}[h]
	\centering
	\includegraphics[scale=0.45]{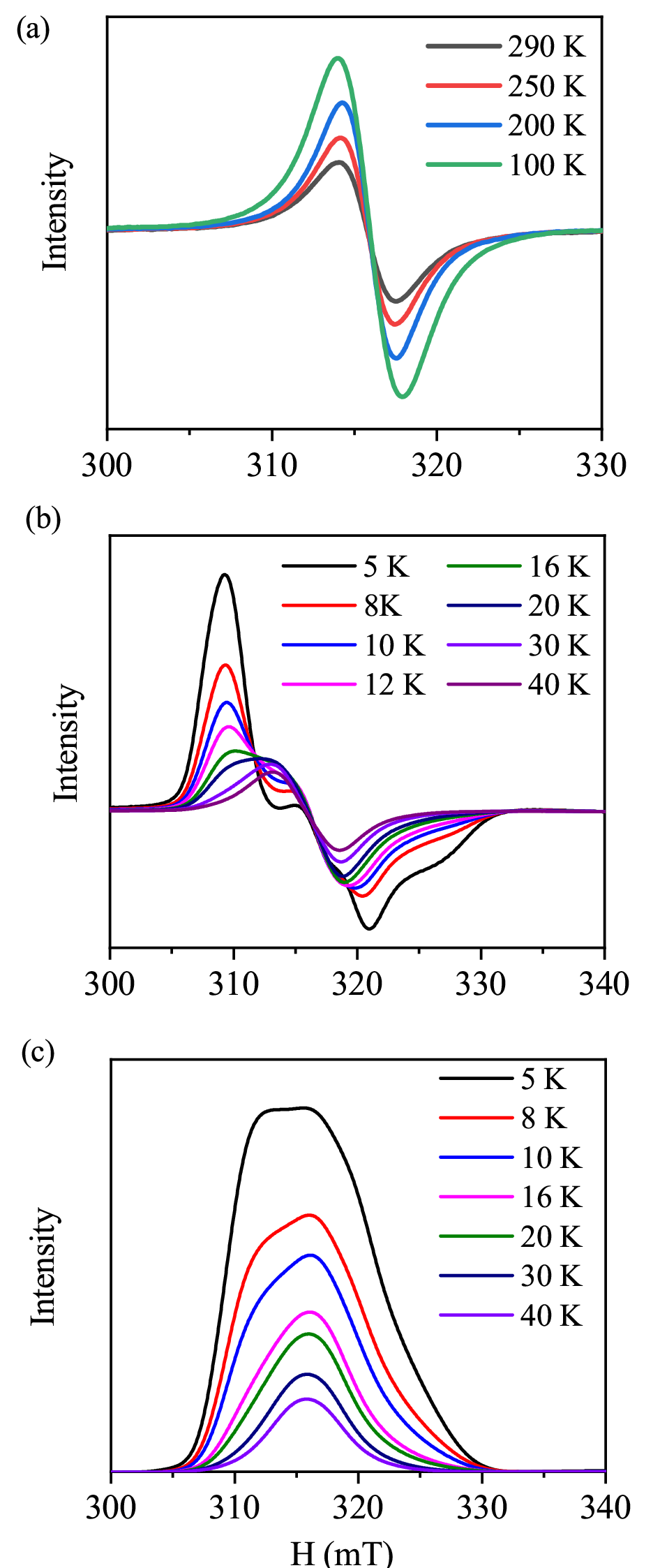}
	\caption{(a) It shows a few representative EPR derivative spectra of a polycrystalline pellet of LCO recorded in the temperature range between T= 300 - 100 K. (b) shows a few derivative ESR spectra at lower temperatures where multiple ESR absorption lines are visible. (c) It shows the integrated spectra of the low-temperature EPR signal shown in Fig. b}
\end{figure}

\begin{figure}[h]
	\centering
	\includegraphics[scale=0.46]{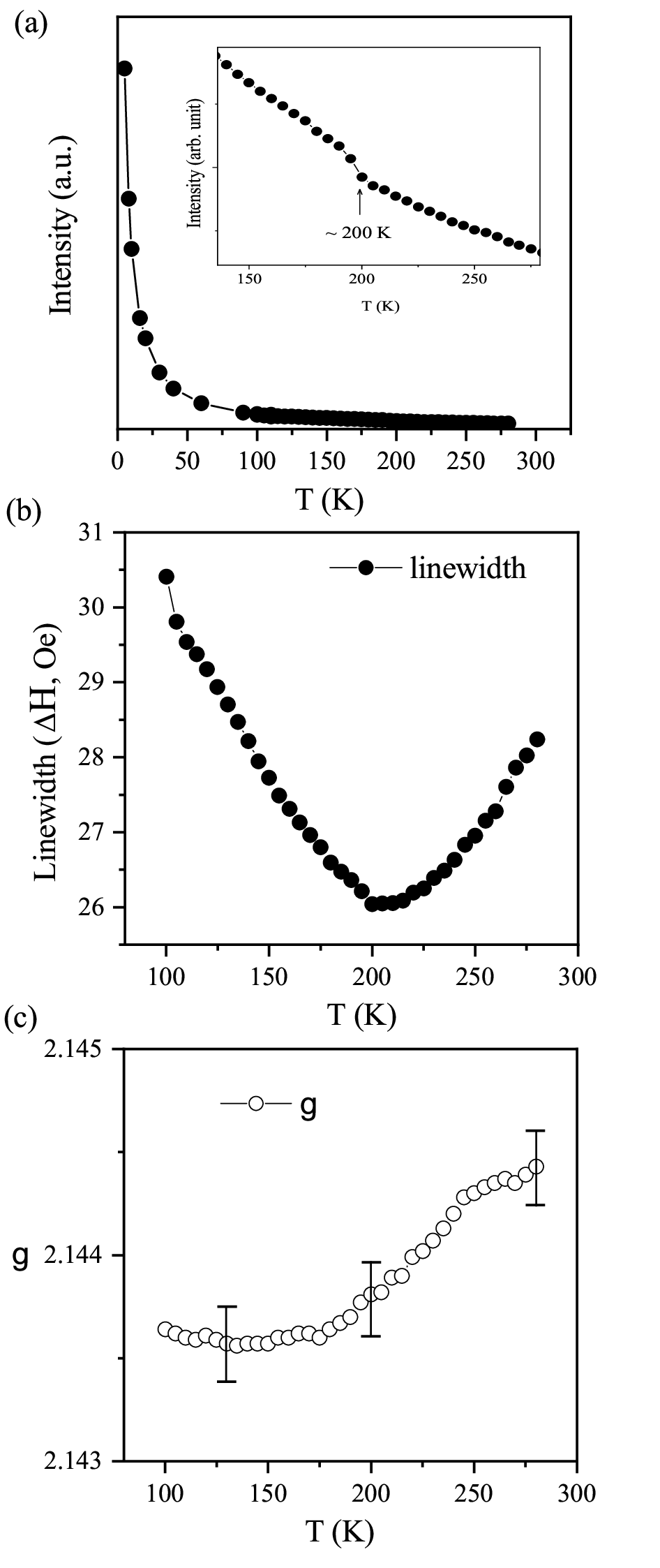}
	\caption{The Fig. shows the parameters (a) intensity, (b) linewidth (c) $g$-factor obtained by fitting the EPR spectra in the temperature range between T = 300 - 100 K, assuming a Lorentzian lineshape.  }
\end{figure}

\subsection{Electron paramagnetic resonance spectroscopy}
EPR (Electron Paramagnetic Resonance) is a useful tool for investigating magnetic systems lacking long-range magnetic order, especially those with local moments, spin dynamics, {\color{blue}\cite{Pal2024, Pal2025}} and should be helpful to understand the magnetic state of the bulk Li$_x$CoO$_2$. Earlier experimental results of EPR investigations on this compound is non-trivial. First of all, the origin of the well-defined and rather sharp EPR signal observed at room temperature is surprising and is not fully understood. The EPR absorption observed in this compound in most cases was related to surface defects, or impurity spins originating from other phases like Co$_3$O$_4$ or CoO. In the stoichiometric LiCoO$_2$, the EPR signal has been associated with thermally activated spin fluctuations {\color{blue}\cite{Kazuhiko2014}}. Similar to previous studies, we also have observed a sharp EPR signal at room temperature with a g-factor and linewidth of 2.12 and 28 $G$, respectively. First of all, such a narrow EPR with a Lorentzian lineshape at $g$ = 2.14 is surprising. In Li$_x$CoO$_2$, Co$^{3+}$ does not have any unpaired electron and therefore should be EPR silent. On the other hand, in compounds with Co$^{4+}$, the EPR signal at room temperature is barely noticeable. For example, in Co$_3$O$_4$, the ESR spectra is visible only at low temperatures and also the linewidth is extremely large. However, the $g$ = 2.14 matches well with the expected $g$ value for Co$^{4+}$ ion. The most important point we want to underline is that, by comparing the relative intensity of the EPR signal of Li$_{x}$CoO$_2$ (x $\sim$ 0.9) with standard sample of CuSO$_4$, 5H$_2$O, we found that only approximately 1\% of the total holes is responsible for the observed EPR signal. This inevitably means that most of the localized spins in the system are EPR silent at the X-band frequency. 

In Fig. {\color{blue}9(a,b)} we show the derivative-EPR spectra at different temperatures. The signal above $T\sim$ 100 K can be modeled assuming a single Lorentzian lineshape. However, at lower temperatures, the spectra undergoes a gradual distortion, which can not be satisfactorily modeled by a single absorption line. Finally, below $T\sim$ 40 K, additional lines are clearly visible as shown in Fig {\color{blue}9(b)}.  At further lower temperatures, approximately $T <$ 25 K, additional ESR lines are clearly visible and becomes prominent at further lower temperatures, resulting into a complex ESR absorption at the lowest measured temperature of $T$ = 5 K. In Fig. {\color{blue}9(c)}, we show integrated spectra at low temperatures to further highlight the multiple EPR absorption. The temperature dependence of the EPR spectra and the obtained parameters by fitting each spectra provide further insightful information about the possible source of the EPR signal {\color{blue}\cite{Sichelschmidt1995,Loidl1997,Kochelaev1997, Matsui2008}}. In Fig. {\color{blue}10(a-c)}, we have shown the temperature dependence of the EPR intensity down to 5 K, and linewidth and $g$-factor down to 100 K. To estimate the intensity, we have double integrated the spectra at individual temperatures. We see that the intensity, which is proportional to the static susceptibility of EPR centers, slowly increases at higher temperatures and increases rapidly at lower temperatures. Interestingly, we can identify a temperature scale $T_{min}\sim$ 200 K, where the intensity shows a weak anomaly. This matches the temperature where a broad minima is observed in $ZFC$ curve at low fields. The EPR linewidth gradually reduces, goes through a minima around $T_{min}\sim$ 200 K and increases quite linearly at further lower temperatures. The $g$-factor slowly decreases and appears to achieve a stable value similar temperature scale. We propose that EPR signal appears from isolated small polarons. Above $T_{min} \sim$ 200 K, there is a thermally activated slow hopping of the polarons, which results into increasing EPR linewidth with increasing temperature {\color{blue}\cite{Shengelaya1996, Shengelaya2000}}. Below $T_{min}$, the hopping gradually freezes and the linewdith increases again with decreasing temperature. At very low temperature, we finally see the powder pattern arising from a localised polarons with anisotropic $g$-tensor. 

\section{Discussion:}
In the present compound, Li-ion deficiency introduces holes in the system. Previous soft X-ray spectroscopy measurements revealed that the localized holes are bound to the Co$^{4+}$O$_6$ distorted octahedra {\color{blue}\cite{Mizokawa2013}}, and suggested formation of polarons. First-principle calculation further proposed ferromagnetically or anti-ferromagnetically coupled polarons {\color{blue}\cite{Ahn2023}}. The localized  holes distort it's surrounding CoO$_6$ octahedra to form small polarons and reside close to the Li vacancies {\color{blue}\cite{Mizokawa2013,Ahn2023,Menetrier1999}}. Now based on our bulk magnetization, heat capacity and electron paramagnetic resonance spectroscopy results, we propose that the small polarons give rise to short-range ferromagnetic interaction in small clusters of doped holes which in turn giving rise to history dependence in magnetization measurements even at room temperatures. In lightly hole doped LaCoO$_3$, where Co$^{3+}$ remains in low spin state at low temperatures and hence is non-magnetic and insulator. However, replacing La$^{3+}$ by Sr$^{2+}$ or Ca$^{2+}$ results into magnetic polaron formation and hole-rich ferromagnetic clusters {\color{blue}\cite{Kuhns2003,1Phelan2006,Goodenough1995,Phelan2006,Khomskii2008,Kataev2010,Khomskii2011}}. In our case, the high-temperature magnetic state of Li$_{0.9}$CoO$_2$ comprises of a kind of mixed state containing both polaron clusters with short-range ferromagnetic interaction and isolated holes at CoO$_{6}$ octahedra. At low temperatures, some fraction of the hole clusters also undergo an antiferromagnetic transition below $T_N \sim$ 10 K. Nevertheless, further experiments are necessary to completely understand the microscopic details. Comparing with hole-doped LaCoO$_3$, our study indicates towards a general trend in lightly hole-doped Co-based oxides.    

In this context, it may be worthwhile to investigate Na$_x$CoO$_2$ which also shows superconductivity. Its due to the uncanny resemblance of the temperature dependence of EPR parameters with high-T$_c$ cuprates. Particularly, the distinct minima that is observed in the linewidth {\color{blue}Fig. 7(b)}, closely resembles the temperature dependence of the EPR linewidth  in cuprates, where one hole at oxygen and two holes in adjacent Cu forms a quasi-particle, called as the three-spin-polarons (TSP) {\color{blue}\cite{Sichelschmidt1995,Kochelaev1997}}. In high-T$_c$ cuprate superconductor, Emery and Reiter proposed a model for the superconductivity in terms of the so-called three-spin-polaron. This forms out of an oxygen hole coupled to two neighboring copper holes {\color{blue}\cite{Emery1988}}. The ground state of such three-spin-polarons were found to be a spin doublet {\color{blue}\cite{Kochelaev2003}}. There is growing evidence that polaron formation and mutual interaction between polarons play an essential role in the spin and charge dynamics in these materials and the high-Tc superconductivity as well. In these compounds, EPR has provided important information on the quasi-particles. At high temperatures, the TSP can move through the CoO$_2$ layer by tunneling processes as Li ions diffuse through the vacancies. The concentration of the polarons should depend on the concentration of the Li-ion vacancies. In addition, the grain boundaries may also affect the motion. Importantly, at high enough local concentration of the polarons, they may merge to give rise to macroscopic clusters or drops with local ferromagnetic correlation {\color{blue}\cite{Bersuker1997}}, which will be EPR silent and hence consistent with the small amount of EPR active spins in the system.  

\section{\label{sec:level}Summary:}
In summary, we have prepared polycrystalline sample of Li deficient Li$_x$CoO$_2$ (x$\sim$0.9) and measured temperature dependence of the bulk magnetization, heat capacity, and electron paramagnetic resonance spectroscopy to understand its magnetic property. We found that this compound shows history-dependent magnetization at low fields in a wide temperature range and a weak ferromagnetism even at room temperature. In addition, magnetization and heat capacity shows a broad and weak hump possibly due to partial antiferromagnetic order below $T_N \sim$ 10 K. We propose that the magnetic response of this compound primarily driven by the formation of magnetic polarons which is intricately linked to Li diffusion at high temperatures. The history dependence of magnetization resembling weak ferromagnetism and temperature-dependent EPR spectroscopy suggest the three-spin-polarons at higher temperatures similar to high-temperature superconductors.

\section{\label{sec:level} Acknowledgment:}
We acknowledge Eva Brucher for help during heat capacity and electron spin resonance spectroscopy measurements. We thank Dr. K. Von Nidda and Dr. Rajeev Rawat for discussion. We acknowledge Dr. Archana Sagdeo, RRCAT for the synchrotron XRD measurements, and Dr. N. P. Lalla for the X-ray diffraction measurements using Cu-K$_\alpha$. We acknowledge Mr. Samir Hammoud at Max Planck Institute for solid state physics, Stuttgart, Germany for his help in ICP-OES and hot gas extraction measurements required for the determination of the stoichiometry of the investigated compound. 
Research at TU Wien was supported by HEC Pakistan (HEC/HRD/OSS-III/Batch-3/Austria/2025/1773; administered by OeAD).

\end{document}